\documentstyle[12pt,epsf]{article}
\renewcommand{\bar}[1]{\overline{#1}}

\renewcommand{\d}{{\mathrm d}}

\begin{document}

\begin{flushright}
USM-TH-113 
\end{flushright}
\bigskip\bigskip


\centerline{\large \bf Azimuthal Spin Asymmetries of Pion Electroproduction}

\vspace{18pt} \centerline{\bf Bo-Qiang Ma
$^{a}$,
Ivan Schmidt
$^{b}$,
and Jian-Jun Yang
$^{b,c}$}

\vspace{8pt}

{\centerline {$^{a}$Department of Physics, Peking University,
Beijing 100871, China,}}

{\centerline {and CCAST (World Laboratory),
P.O.~Box 8730, Beijing 100080, China}}


{\centerline {$^{b}$Departamento de F\'\i sica, Universidad
T\'ecnica Federico Santa Mar\'\i a,}}

{\centerline {Casilla 110-V, 
Valpara\'\i so, Chile}}

{\centerline {$^{c}$Department of Physics, Nanjing Normal
University, Nanjing 210097, China}}

\vspace{6pt}
\begin{center} {\large \bf Abstract}

\end{center}
Azimuthal spin asymmetries, both for charged and neutral pion
production in semi-inclusive deep inelastic scattering of
unpolarized charged lepton beams on  longitudinally and
transversely polarized nucleon targets, are analyzed and
calculated. Various assumptions and approximations in the quark
distributions and fragmentation functions often used in these calculations
are  studied in detail. It is found that different approaches to the
distribution  and fragmentation functions may lead to quite
different predictions on the azimuthal asymmetries measured in the
HERMES experiments, thus their effects should be taken into
account before using the available data as a measurement of quark
transversity distributions. It is also found that the unfavored
quark to pion fragmentation functions must be taken into account
for $\pi^-$ production from a proton target, although they can be
neglected for $\pi^+$ and $\pi^0$ production. Pion production
from a proton target is suitable to study the $u$ quark
transversity distribution, whereas a combination of pion
production from both  proton and neutron targets can measure
the flavor structure of quark transversity distributions.

\vfill

\centerline{PACS numbers: 13.87.Fh, 13.60.-r, 13.88.+e,
14.20.Dh}

\vfill

\newpage

\section{Introduction}

Recently, the HERMES collaboration reported the observation of
single-spin azimuthal asymmetries for charged and neutral pion
production, in semi-inclusive deep inelastic scattering (DIS) of
unpolarized positron beam on the longitudinally polarized proton
target \cite{HERMES00,HERMES01}. Such azimuthal asymmetries are 
important, because they can provide information of the
transversity distributions \cite{TransversityReview}, which are
one of the three fundamental quark distributions of the nucleon.
The other two fundamental quark distributions are the momentum and
helicity distributions, which have been explored for more than
three decades in various DIS processes
\cite{DISprocess,DYprocess}, and their explicit flavor-dependence
has been obtained  with  relative high precision, although there are still
some uncertainties concerning the sea content at small Bjorken
variable $x$ and the valence flavor-spin structure as $x \to
1$. However, the experimental measurements of the quark
transversity distributions are just about to begin, since the
transversity is not directly observable in inclusive DIS
processes. It has been proposed that the transversity can manifest
itself through the Collins effect \cite{Col93} of nonzero
production between a chiral-odd structure function and a T-odd
fragmentation function, which is accessible in some specific
semi-inclusive hadron production experiments 
\cite{Col93,Kot95,Ans95,Mul96,Kot97,Jaf98,Ams}. Indeed, there have
been a number of studies
\cite{Kot99,Efr00,Sch00,Bor99,Bog00,Ans00,Suz00,MSY9,Oga98,Oga00,Kor00,Efr01} to
show that the azimuthal asymmetries measured by HERMES can be used
to extract the quark transversity distributions of the nucleon,
although the theoretical predictions of the explicit
flavor-dependent transversity distributions are different in various
models \cite{TransversityReview}. Therefore the HERMES
measurements provide us the opportunity to confront theoretical
predictions with physical observations, and it is meaningful to
examine  the theoretical results carefully, in
order to see what conclusion we can draw on the transversity.

The analyzing power of azimuthal spin asymmetry measured by HERMES is defined as
\begin{equation}
A_{UL}^W=\frac{\int \left[{\mathrm d} \phi\right] W(\phi)
\left\{N^+(\phi)-N^-(\phi)\right\}}{\frac{1}{2}\int \left[{\mathrm
d} \phi\right]\left\{ N^+(\phi)+N^-(\phi) \right\}}, \label{AP}
\end{equation}
where $UL$ denotes {\it unpolarized} beam on a
{\it longitudinally} polarized target, $W(\phi)=\sin \phi$ or $\sin 2
\phi$ is the weighting function for picking up the Collins effect,
and $N^+(\phi)$ ($N^-(\phi)$) is the number of events for pion
production, as a function of $\phi$, when the target is positively
(negatively) polarized. The azimuthal
angle $\phi$ is the angle between the pion emitting plane
and the lepton scattering plane, with the lepton scattering plane determined
by the incident
and scattered leptons, and the pion emitting plane determined
by the final detected pion and the virtual photon. The
virtual photon acts as the common axis of both planes.
The analyzing powers of azimuthal asymmetries for charged pions, $\pi^+$ and
$\pi^-$, and neutral pion, $\pi^0$, have been measured \cite{HERMES00,HERMES01},
and there is clear evidence
for  non-zero values of $A_{UL}^{\sin \phi}$ for $\pi^+$ and $\pi^0$ 
production, which
indicates the existence of azimuthal asymmetries. Under a number of simplifying assumptions and
approximations, the analyzing power $A_{UL}^{\sin \phi}$ of azimuthal asymmetry
is proportional
to the ratio
\begin{equation}
\frac{\sum_q e^2_q \delta q (x) H_{1}^{\perp(1)q}(z)
}
{\sum_q e^2_q q(x) 
D_{1}^{q}(z)
}.
\label{AP1}
\end{equation}
Here $e_q$ is the charge of the quark with flavor $q$, $q(x)$
and $\delta q(x) $ are the quark momentum and transversity
distributions of the nucleon target, $D_{1}^{q}(z)$ is the
fragmentation function  for an unpolarized quark
with flavor $a$ into a pion $\pi$, and $H_{1}^{\perp(1)q}(z)$ is 
the so called Collins function describing the fragmentation of 
a transversely polarized quark into a pion $\pi$. If one assumes, as adopted in
almost all previous analysis, that the pion production is
dominated by the ``favored" fragmentation processes $ u \to \pi^+$
and $d \to \pi^-$ for charged pions, and neglects the ``unfavored"
processes $ u \to \pi^-$ and $d \to \pi^+$, one may conclude that
the $\pi^+$ azimuthal asymmetry measures the $u$-quark
transversity $\delta u(x)$ and the $\pi^-$ azimuthal asymmetry
measures the $d$-quark transversity $\delta d(x)$, respectively.
This is the situation of the current available studies.

The purpose of this paper is to take into account the terms neglected
in (\ref{AP1}), and also to take into consideration  the contribution from
the ``unfavored" fragmentation process. We will estimate their
effects on the azimuthal asymmetries for the three kinds of produced pions:
$\pi^+$, $\pi^-$, and $\pi^0$. We find that the neglected terms
may have sizable effects on the azimuthal asymmetries in the
HERMES kinematical region, and that the results are sensitive to
different assumptions on the quark distributions and fragmentation
functions. We find also that, although the neglect of the
``unfavored" fragmentation processes seems to be reasonable for
$\pi^+$ production, it is not justified for $\pi^-$ production.
Moreover, a significant difference between the $\pi^-$ azimuthal
asymmetries predicted by two different models \cite{MSY9}, are
largely reduced by taking into account the ``unfavored" processes, and this
renders it difficult to use $\pi^-$ production as a clean process
to test different predictions on the flavor-spin quark structure
of the nucleon at large $x$. We will also check the possible
ambiguities from the uncertainties in the ``unfavored"
fragmentation functions. However, it will be shown that the
favored fragmentation process in $\pi^+$ production from a
neutron target plays a more important role than that in the
corresponding $\pi^-$ production from a proton target. Thus 
$\pi^+$ production from a neutron target can provide a test of
different model predictions. A combination of pion production from
both  proton and neutron targets can measure both of the $u$
and $d$ quark transversity distributions,  $\delta u(x)$ and
$\delta d(x)$, of the proton.

This paper is organized as follows. In Section II of the paper, we
will present the formulae for the azimuthal asymmetries, and
analyze in details various assumptions and approximations on
the quark distributions and fragmentation functions used for
numerical calculations. In section III, we perform numerical
calculations of the azimuthal asymmetries for pions:
$\pi^+$, $\pi^-$, and $\pi^0$, and give detailed discussions on
the effects of using  different approaches to the  distribution functions
and fragmentation functions. We discuss also in detail
the uncertainties  coming from the ``unfavored" processes of
quark to pion fragmentation. In Section IV, we predict the
azimuthal asymmetries for pion production in semi-inclusive DIS
processes of an unpolarized charged lepton beam on a transversely
polarized nucleon target. Finally, we present the summary and
conclusions in Section V.

\section{Formalism of Azimuthal Spin Asymmetry in Pion Electroproduction}

Deep inelastic scattering (DIS) of a lepton on a nucleon is
one of the most efficient and clean tools to detect the underlying
structure of the nucleon.  With the recent progress in 
experimental techniques, the precision detection of hadrons 
produced  in DIS processes makes it
possible to reveal more about the basic structure of the nucleon,
and the use of polarized beams and/or polarized targets makes it
also possible  to measure with some precision various polarized quark
distributions and fragmentation functions. The HERMES experiments
\cite{HERMES00,HERMES01} are performed with an unpolarized lepton
beam scattering on a longitudinally polarized proton target, and
some assumptions and approximations are needed in order to connect
the pion production with the transversity distribution, which is
supposed to be related to the probability of finding a quark with
its spin aligned to the proton spin for a transversity polarized
proton. We first give the formulae for the analyzing power of
azimuthal asymmetry and analyze the quantities needed to perform
numerical calculations.

\subsection{Spin asymmetries of pion electroproduction}

We now introduce the basic kinematics for semi-inclusive DIS
processes. Let $l_1$ and $l_2$ be the initial and final momenta of
the incoming and outgoing charged leptons respectively, $Q^2=-q^2$ be the
squared 4-momentum transfer of the virtual photon with momentum
$q=l_1-l_2$, and $P$ and $P_h$ ($M$ and $M_h$) be the target and final
hadron momenta (masses). Then the three basic variables, $x$, $y$,
and $z$, used in describing DIS processes can be expressed as
\begin{equation}
x=\frac{Q^2}{2 P \cdot q};  ~~~~
y=\frac{P \cdot q}{P \cdot l_1}; ~~~~
z=\frac{P \cdot P_h}{P \cdot q},
\end{equation}
where $x$ is  the Bjorken variable which corresponds to the
momentum fraction carried by the struck quark in a light-cone
description, $y$ is the fraction of the initial lepton's energy
transferred to the quark by the virtual photon, and $z$ is the
fraction of quark momentum transferred to the final produced
hadron in the fragmentation process. In the target rest frame, the
three variables can be expressed as
\begin{equation}
x=\frac{Q^2}{2 M \nu};  ~~~~
y=\frac{\nu}{E}; ~~~~
z=\frac{E_{h}}{\nu},
\end{equation}
where $\nu$, $E$, and $E_{h}$ are the virtual photon energy,
the incident lepton energy,
and the produced hadron energy, respectively.
$s=2 P \cdot l= 2 M E$ is the total squared energy
in the lepton-proton center of mass frame,
therefore we have a relation $Q^2=s\, x \,y$. The incident
lepton energy is $E=27.6$ GeV in the HERMES experiments.
The invariant mass of the photon-proton
system is $W=\sqrt{2M \nu +M^2-Q^2}$.
Let $P_{h\perp}$ be the final hadron transverse momentum and $l_{1\perp}$ be the
initial lepton transverse momentum, 
both respect to  the virtual photon momentum
direction. Then the azimuthal angle $\phi$ is the angle between $P_{h\perp}$ and
$l_{1\perp}$ around the virtual photon direction.

Using the above measurable variables of semi-inclusive hadron production and
the measurable cross sections, one can defined $\sin \phi$ and $\sin 2 \phi$ weighted
spin asymmetries
\begin{equation}
\left< \frac{|P_{h\perp}|}{M_{h}} \sin \phi \right>
=\frac{\int \d ^2 P_{h\perp} \frac{|P_{h\perp}|}{M_{h}} \sin \phi (\d \sigma^+ - \d \sigma^- ) }
{\int d ^2 P_{h\perp} ( \d \sigma^+ + \d \sigma^- ) },
\end{equation}
\begin{equation}
\left< \frac{|P_{h\perp}|^2}{ M\,M_{h}} \sin  2 \phi \right>
=\frac{\int \d ^2 P_{hT} \frac{|P_{hT}|^2}{M\,M_{h}} \sin 2 \phi (\d \sigma^+ - \d \sigma^- ) }
{\int \d ^2 P_{h\perp} ( \d \sigma^+ + \d \sigma^- ) },
\end{equation}
which are measurable quantities. Here +(-) denotes parallel (antiparallel) longitudinal
polarization of the target.
The above asymmetries are related to the azimuthal asymmetries
measured by HERMES through the following relations
\begin{equation}
A_{UL}^{\sin \phi } \approx \frac{2 M_h}{\left< P_{h\perp}\right>}
\left< \frac{|P_{h\perp}|}{M_{h}} \sin \phi \right>,
\end{equation}
\begin{equation}
A_{UL}^{\sin 2 \phi } \approx \frac{2 M\,M_h}{\left< P^2_{h\perp}\right>}
\left< \frac{|P_{h\perp}|^2}{ M\,M_{h}} \sin  2 \phi \right>.
\end{equation}
In the HERMES experiments, $\left< P_{h\perp}\right>=0.44$ GeV, and we may use
$\left< P^2_{h\perp}\right>=(0.44)^2$ GeV$^2$ as an estimate of $\left< P^2_{h\perp}\right>$ in our
calculations.

From another side, one can use the quark-parton model to calculate
hadron production in semi-inclusive DIS processes, and express the
cross sections in terms of various distribution  and
fragmentation functions. Then one can define the weighted cross
section for one flavor quark, $d \sigma_q$, with different
weighting functions depending on the final hadron transverse
momenta $w_i(P_{h\perp})$:
\begin{equation}
\Sigma_i^q=\int \d ^2 P_{h\perp}\, w_i(P_{h\perp}) \d \sigma_q,
\end{equation}
where $w_i(P_{h\perp})=1$, $w_i(P_{h\perp})=P_{h\perp} \sin \phi
/M_h$, and $w_i(P_{h\perp})=|P_{h\perp}|^2 \sin 2 \phi /M M_h$.
Thus one can relate the theoretical calculations with the above
spin asymmetries by
\begin{equation}
\left< \frac{|P_{h\perp}|}{M_{h}} \sin \phi \right>
=\frac{\Sigma_2}{\Sigma_1},
\label{SA1phi}
\end{equation}
\begin{equation}
\left< \frac{|P_{h\perp}|^2}{ M\,M_{h}} \sin  2 \phi \right>
=\frac{\Sigma_3}{\Sigma_1},
\label{SA2phi}
\end{equation}
where a sum over all quark flavors, $\Sigma_i=\sum_q e_q^2 \Sigma_i^q$, 
is implicitly assumed, and will be assumed from now on. In the case of unpolarized
beam and longitudinal polarized target, $\Sigma_1$, $\Sigma_2=\Sigma_{2L}+\Sigma_{2T}$, and
$\Sigma_3$ are given by \cite{Kot95,Mul96,Oga98,Oga00}
\begin{equation}
\Sigma_1=\left[1+(1-y)^2\right]f_1(x) D_1(z),
\end{equation}
\begin{equation}
\Sigma_{2L}=4 S_L \frac{M}{Q}(2-y)\sqrt{1-y} \left[
x \, h_L(x) z  H_1^{\perp (1)}(z) - h_{1L}^{\perp (1)}(x) \tilde{H}(z)\right],
\end{equation}
\begin{equation}
\Sigma_{2T}=2 S_{T x}(1-y) h_1(x) z H_1^{\perp (1)}(z) ,
\end{equation}
\begin{equation}
\Sigma_{3}=8 S_{L}(1-y) h_{1L}^{\perp (1) }(x) z^2 H_1^{\perp (1)}(z).
\end{equation}
Here the components of the longitudinal and transverse target polarization in the virtual
photon frame are denoted by $S_L$ and $S_{T x}$, respectively:
\begin{equation}
S_L=S \cos \theta_{\gamma}, ~~~~ S_{Tx}=S \sin \theta_{\gamma},
\end{equation}
with
\begin{equation}
\sin \theta_{\gamma}=\sqrt{\frac{4M^2x^2}{Q^2+4M^2x^2}\left(1-y-\frac{M^2 x^2y^2}{Q^2}\right)}
=\sqrt{\frac{4M^2x}{s\, y+4M^2x}\left(1-y-\frac{M^2 x\,y}{s}\right)},
\end{equation}
and $S$ is the target polarization, which has the value 0.86 in the HERMES experiments.
Twist-2 distribution functions and fragmentation functions have a subscribe ``1": $f_1$ and $D_1$
are the usual unpolarized distribution and fragmentation function, while  $h_{1L}^{\perp (1)}(x)$
and $h_1(x)$ are the quark transverse spin distribution functions of longitudinally and transversely
polarized nucleons, respectively. $h_L(x)$ is the twist-3 
distribution function of a longitudinally
polarized nucleon, and it can be split into a twist-2 part, $h_{1L}^{\perp (1)}(x)$, and an interaction
dependent part, $\tilde{h}_L(x)$:
\begin{equation}
h_L(x)=-2 \frac{h_{1L}^{\perp (1)}(x)}{x}+\tilde{h}_L(x).
\label{HLL}
\end{equation}
The fragmentation function $\tilde{H}(z)$ is the interaction dependent part of the twist-3 fragmentation function:
\begin{equation}
H(z)=-2 z H_1^{\perp (1)}(z)+\tilde{H}(z).
\end{equation}
The functions with superscript ``(1)" denote $p_{\perp}^2$- and $k_{\perp}^2$-moments, respectively:
\begin{equation}
h_{1L}^{\perp (1) }(x) \equiv \int \d^2 p_{\perp} \frac{p_{\perp}^2}{2M^2}h_{1L}^{\perp }(x,p_{\perp}^2 ),
\end{equation}
\begin{equation}
H_{1L}^{\perp (1) }(z) \equiv  z^2 \int \d^2 k_{\perp} \frac{k_{\perp}^2}{2M_h^2}H_{1L}^{\perp }(z,z^2k_{\perp}^2 ),
\end{equation}
where $p_{\perp}$ and $k_{\perp}$ are the intrinsic transverse momenta of the initial and final partons
in the target and produced hadrons, respectively.

To calculate the spin asymmetries and compare them with
experiments, we need to know the quark distribution functions:
$f_1(x)$, $h_1(x)$, $\tilde{h}_L(x)$, and $h_{1L}^{\perp (1)
}(x)$, and the fragmentation functions: $D_1(z)$, $H_1^{\perp
(1)}(z)$, and $\tilde{H}(z)$. Unfortunately, most of the
distribution functions and fragmentation functions which appear in
these expressions are not known a priori, since they have not been
measured yet. Thus we have to make some assumptions and
approximations, and this leads to different approaches
for the distribution functions and fragmentation functions
\cite{Bog00,Oga00}.

\subsection{Different approaches to the  distribution functions and fragmentation functions}

The most simple approach, denoted as Leading Approach, is to neglect the
$1/Q$ term $\Sigma_{2L}$ in $\Sigma_{2}$, i.e., we neglect both the
${\tilde h}_L(x)$ and $h_{1L}^{\perp (1) }(x)$ terms in the spin asymmetry
$\left< \frac{|P_{h\perp}|}{M_{h}} \sin \phi \right>$. Then we find immediately
that
\begin{equation}
\left< \frac{|P_{h\perp}|}{M_{h}} \sin \phi \right>
=\frac{\Sigma_{2T}}{\Sigma_1}
\propto \frac{h_1(x) H_1^{\perp (1)}(z) }{f_1(x) D_1(z)},
\end{equation}
which is Eq.~(\ref{AP1}). Thus we can relate the quark
transversity distribution $h_1(x)$ to the spin asymmetry with its
$x$-dependence. This is the situation discussed in several
papers \cite{Efr00,MSY9}.

The next approach, denoted as Approach 1, is to assume that the
twist-2 quark transverse spin distribution function of
longitudinally polarized nucleon, $h_{1L}^{\perp (1) }(x)$, is
zero \cite{Oga00}. Then it follows that
\begin{equation}
h_L(x)=\tilde{h}_L(x)=h_1(x).
\end{equation}

Another approach, denoted as Approach 2, is to assume that the
interaction dependent twist-3 part, $\tilde{h}_L(x)$, 
is zero \cite{Kot99}, and
consequently, we can also assume that $\tilde{H}(z)$ is zero. Then
by neglecting the term proportional to the current quark mass, one
can obtain a Wandzura-Wilczek type relation \cite{jaffe92,tan94}
\begin{equation}
h_{1L}^{\perp (1) }(x)=-x^2 \int_x^1 \d \xi \frac{h_1(\xi)}{\xi^2}.
\end{equation}
It follows, from Eq.~(\ref{HLL}), that
\begin{equation}
h_{L}(x)=2x \int_x^1 \d \xi \frac{h_1(\xi)}{\xi^2}.
\end{equation}

In the above two approaches, we only need the distribution functions, $f_1(x)$ and $h_1(x)$,
and the fragmentation functions, $D_1(z)$ and $H_1^{\perp (1)}(z)$, 
in order to calculate the
spin asymmetries $\left< \frac{|P_{h\perp}|}{M_{h}} \sin \phi \right>$ and
$\left< \frac{|P_{h\perp}|^2}{ M\,M_{h}} \sin  2 \phi \right>$.

\subsection{Quark transversity distributions of the nucleon}

In  the three approaches described above, we
need the quark distribution functions, $f_1(x)=q(x)$ and
$h_1(x)=\delta q(x)$, as inputs. The $x$-dependence of the spin
asymmetries are controlled by the above two quark distribution functions,
in which the unpolarized quark distributions $q(x)$ are
known to relative  high precision, thus we are able to extract the
information on the quark transversity distributions $\delta q(x)$
by comparing theoretical predictions with the experimental data on
spin asymmetries.

Although the quark momentum distributions $q(x)$ have been measured to 
high precision, there are still uncertainties concerning
the flavor structure as $x \to 1$. For example, the SU(6)
quark spectator diquark model \cite{Fey72,DQM,Ma96} predicts
$\left. d(x)/u(x) \right|_{x=1}=0$, whereas a pQCD based counting
rule analysis \cite{Far75,countingr,Bro95} predicts $\left.
d(x)/u(x) \right|_{x=1}=1/5$. The flavor structure of the
transversity distributions in the two models are also found to
differ significantly in this region\cite{MSY9,MSSY8}: the pQCD based analysis
predicts \cite{MSY9,MSSY8} $\delta d(x)/ d(x) \to 1$, whereas the
SU(6) quark-spectator-diquark model predicts \cite{Ma98a,Ma98b}
$\delta d(x)/ d(x) \to -1/3$. Detailed constructions of the quark
transversity distributions in the two models can be found in
Refs.~\cite{MSY9,MSSY8}.

To make realistic predictions of measurable quantities, we need also
to take into account the sea quark contribution in the two model constructions.
In the quark diquark model case, this can be achieved by adopting one set of unpolarized
quark distribution parametrization as input, and then use theoretical relations
to connect the quark transversity distributions
with the unpolarized quark distributions \cite{Ma96,Ma98a}:
\begin{equation}
\begin{array}{clcr}
\delta u_{v}(x)
    =[u_v(x)-\frac{1}{2}d_v(x)]\hat{W}_S(x)-\frac{1}{6}d_v(x)\hat{W}_V(x);\\
\delta d_{v}(x)=-\frac{1}{3}d_v(x)\hat{W}_V(x), \label{udv}
\end{array}
\end{equation}
in a similar way as was done for the quark helicity distributions
\cite{Ma96}. $\hat{W}_S(x)$ and $\hat{W}_S(x)$, which come from the
relativistic effect of quark transversal motions \cite{Ma91b},
are the Melosh-Wigner rotation factors \cite{Ma98a,Ma98b,Sch97}
for spectator scalar and vector diquarks.
We  use the valence quark momentum distributions
$u_v(x)$ and $d_v(x)$ from quark distribution
parametrization,  but with
$\hat{W}_S(x)$ and $\hat{W}_V(x)$
from the model calculation \cite{Ma98a}. In this way we can
take into account the sea contribution for the unpolarized quark distributions
through the input parametrization. This can provide a more
reliable prediction for the absolute magnitude and shape of a
physical quantity than directly from the model calculation.

For the pQCD based analysis, we adopt the same consideration as above and make
the following connection to relate the pQCD model quark distributions
with the parametrization
\begin{equation}
u_v^{pQCD}(x)=u_v^{para}(x), ~~~~  d_v^{pQCD}(x)=\frac{d_v^{th}(x)}{u_v^{th}(x)} u_v^{para}(x),
\end{equation}
\begin{equation}
\delta u_v^{pQCD}(x)=\frac{\delta u_v^{th}(x)}{u_v^{th}(x)} u_v^{para}(x),   ~~~~
\delta d_v^{pQCD}(x)=\frac{\delta d_v^{th}(x)}{u_v^{th}(x)} u_v^{para}(x),
\end{equation}
where the superscripts ``th" means the pure theoretical
calculation in the pQCD analysis \cite{MSY9,MSSY8}, and ``para"
means the input from parametrization. The superscript ``pQCD"
means that the new quark distributions keep exactly the same
flavor and spin structure as that in the pQCD analysis, but with
detailed $x$-dependent behaviors more closed to the realistic
situation. This is equivalent to use an unique correction factor,
$u_{v}^{para}(x)/u_{v}^{th}$, to adjust each pure {\it
theoretically} calculated quantity to a more realistic pQCD {\it
model} quantity. In this way we can take into account the sea
contribution by using the sea quark distributions from
parametrization, while still keep the pQCD model behaviors of the
valence quark distributions.

\vspace{0.3cm}
\begin{figure}[htb]
\begin{center}
\leavevmode {\epsfysize=3.5cm \epsffile{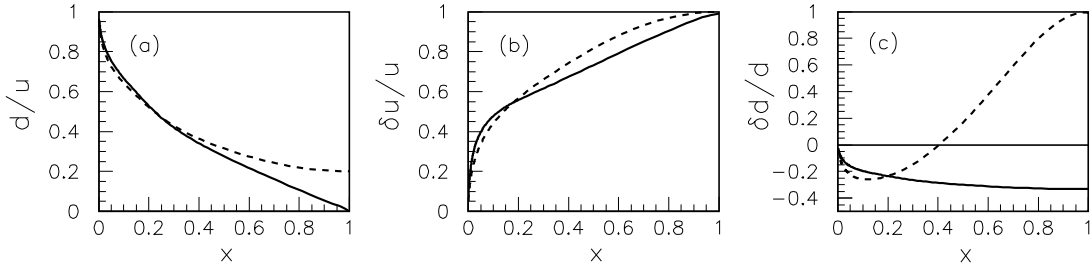}}
\end{center}
\caption[*]{\baselineskip 13pt The different flavor and
transversity structure of the two model quark distributions: (a)
$d_v(x)/u_v(x)$; (b) $\delta u_v(x)/u_v(x)$; and (c) $\delta
d_v(x)/d_v(x)$.  The solid and dashed curves correspond to the
quark-diquark model and the pQCD analysis, respectively.
}\label{msy10fa}
\end{figure}

The explicit flavor and transversity structure of the two sets of
quark distributions can be found in Fig.~\ref{msy10fa}.
Thus we
have two sets of quark distributions of $q(x)$ and $\delta q(x)$,
which keeps the same valence behaviors as in the quark diquark
model and pQCD based model predictions, but with the sea from
parametrization. We expect to make some more realistic predictions
based on these two sets of quark distributions, while still can
distinguish between the different flavor and spin structure
predicted by the two models.

There might be some uncertainties of the magnitude of the explicit
quark distributions, however, these uncertainties will be reduced
since we will only calculate quantities which involve the ratios
of different quark distributions in our later numerical
calculations.

\subsection{Fragmentation functions of pion}

We now turn our attention to the fragmentation functions $D_1(z)$
and  $H_1^{\perp (1)}(z)$. $D_1(z)$ is the unpolarized
fragmentation function for a quark splitting into a pion, and in previous
studies on the azimuthal asymmetries, only the favored
fragmentation functions, e.g., for a $u$ or $\bar{d}$ quark into
$\pi^+$, are considered. However, a new parametrization
of quark to pion fragmentation functions has been proposed \cite{Kre01},
which  combines
pion production data from both semi-inclusive DIS and $e^+e^-$
processes. This new parametrization provides a complete set of
both favored and unfavored fragmentation functions, and therefore it is
very useful for the extraction of nucleon structure information
from pion production in semi-inclusive DIS processes. In this
paper, we will adopt this set of quark to pion fragmentation
functions in our analysis of the transversity distributions
through the azimuthal spin asymmetries discussed above.

For the favored fragmentation function, we mean the fragmentation function
describing the transition of a ``valence" quark into the pion, and it is,
assuming isospin symmetry,
\begin{equation}
D(z)=D_u^{\pi^+}(z)=D_{\bar{d}}^{\pi^+}(z)=D_{d}^{\pi^-}(z)=D_{\bar{u}}^{\pi^-}(z)
\end{equation}
for charged pion fragmentation. The unfavored fragmentation
function is supposed to describe the transition of a light-flavor
``sea" quark into  pions, and it is
\begin{equation}
\hat{D}(z)=D_{\bar{u}}^{\pi^+}(z)=D_{d}^{\pi^+}(z)=D_{\bar{d}}^{\pi^-}(z)=
D_{u}^{\pi^-}(z)
\label{ufFF}
\end{equation}
for $\pi^{\pm}$ fragmentation. For the case of the neutral pion,
assuming that
\begin{equation}
D_{q}^{\pi^0}(z)=\frac{1}{2}\left[D_q^{\pi^+}(z)+D_q^{\pi^-}(z) \right],
\end{equation}
we find
\begin{equation}
D^{\pi^0}(z)=D_u^{\pi^0}(z)=D_{\bar{u}}^{\pi^0}(z)=D_{d}^{\pi^0}(z)=D_{\bar{d}}^{\pi^0}(z)=
\frac{1}{2} \left[D(z)+\hat{D}(z)\right].
\end{equation}
The explicit analytically forms of $D(z)$ and $\hat{D}(z)$ are given by \cite{Kre01}
\begin{equation}
\begin{array}{ll}
D(z)=0.689 z^{-1.039}(1-z)^{1.241},\\
\hat{D}(z)=0.217 z^{-1.805}(1-z)^{2.037}.
\label{DD}
\end{array}
\end{equation}

From another point of view, assuming the Gribov-Lipatov relation
connecting the fragmentation functions and the distribution
functions \cite{GLR,BRV00,Bar00}, we have
\begin{equation}
D_q^{\pi}(z) \propto z q^{\pi}(z),
\end{equation}
where $q_{\pi}(z)$ is the quark distribution of finding a quark $q$ with momentum fraction
$x=z$ in the pion.
We also know from the pQCD based analysis \cite{countingr,Bro95} that the $x \to 1$ behavior
of the quark distribution in the hadron $h$ satisfies the counting rule
\begin{equation}
q^{h}(x) \propto (1-x)^{2 n_{\mathrm s}-1},
\end{equation}
where $n_{\mathrm s}$ means the spectator quark number. For favored quark fragmentation
we notice that
\begin{equation}
D(z) \propto (1-z)
\end{equation}
because there is only one spectator quark, whereas it is
\begin{equation}
\hat{D}(z) \propto (1-z)^5
\end{equation}
for unfavored quark fragmentation because the spectator quark
number $n_{\mathrm s }=3$ in this case. Thus we may assume
\begin{equation}
\hat{D}(z)=(1-z)^4 D(z)
\end{equation}
in this extreme situation. To reflect the possible uncertainties in the parametrization
of fragmentation functions, we also assume three
options for the relation between favored and unfavored fragmentation functions
\begin{equation}
\hat{D}(z)=(1-z)^n D(z),
\end{equation}
with $n=2$, $3$, $4$,  respectively, to compare with the new parametrization (\ref{DD})
described above.

The so called Collins fragmentation function $H_1^{\perp (1)}(z)$, which describes the transition
of a transversely polarized quark into a pion, has not been systematically
measured yet, and it is also
theoretically not well known. However, there has been a so
called Collins parametrization \cite{Col93} of this
fragmentation function, and we will adopt this parametrization in this paper.
It is suggested
that
\begin{equation}
A_C(z,k_{\perp})=\frac{|k_{\perp}|H_1^{\perp q}(z,z^2k^2_{\perp})}{M_h D_1^q(z,z^2k^2_{\perp})}
=\frac{M_C |k_{\perp}|}{M_C^2+|k_{\perp}^2|},
\end{equation}
with $M_C$ being a typical hadronic scale around $0.3 \to 1$ GeV. Assuming a Gaussian
type of the quark transverse momentum dependence in the unpolarized fragmentation
function
\begin{equation}
D_1^q(z,z^2k^2_{\perp})=D_1^q(z) \frac{R^2}{\pi z^2} \exp(-R^2 k_{\perp}^2),
\end{equation}
one obtains
\begin{equation}
H_1^{\perp (1) q}(z)=D_1^q(z) \frac{M_C}{2M_h}
\left(1-M_C^2 R^2 \int_0^{\infty} \d x \frac{\exp(-x)}{x+M_C^2 R^2}\right),
\end{equation}
where $R^2=z^2/\left<P^2_{h\perp}\right>$, and $\left<P^2_{h\perp}\right>=z^2\left<k_{\perp}^2\right>$
is the mean-square momentum the hadron acquires in the
quark fragmentation process. We will set the parameters
$M_C=0.7$ GeV and $\left<P^2_{h\perp}\right>=(0.44)^2$ GeV$^2$ as they are consistent
with the spin asymmetry measured at HERMES \cite{Kor00}.

In fact, this fragmentation function controls the $z$- and
$P_{h\perp}$-dependence of the azimuthal asymmetries, therefore
the $z$- and $P_{h\perp}$-dependence will not be considered in this paper. We will mainly aim at the physics of
the transversity distributions, which controls the $x$-dependence
of the azimuthal asymmetries. 

\section{Numerical Calculations}

With the kinematics of semi-inclusive process and the formulae for
the azimuthal spin asymmetries of pion production, and also with
 the quark distribution functions and fragmentation
functions given in the above section, we can perform numerical
calculations under different assumptions and options. We will
compare the numerical calculations of the azimuthal asymmetries
with the HERMES experimental data \cite{HERMES00,HERMES01}
constrained
by the following experimental cuts
\begin{equation}
1<Q^2 < 15 ~{\mathrm GeV}^2, ~~~W>2 {\mathrm GeV}, ~~~0.2<y<0.85, ~~~
0.2<z<0.7.
\end{equation}
For the $Q^2$ and $W$ used
in the integration over $y$ and $z$, we use the relations
\begin{equation}
Q^2=s \,x\,y, ~~~~ W^2=s\,y(1-x)+M^2,
\end{equation}
with $s=2 M \,E=51.8$ GeV$^2$ in the HERMES experiments.
For the unpolarized
quark distributions, we use the CTEQ 
parametrization \cite{CTEQ5} (CTEQ5 set 1) as  input for both the 
quark diquark model and pQCD model quark distribution functions described in the above section.

\subsection{Different approaches of distribution functions and fragmentation functions}

\begin{figure}[htb]
\begin{center}
\leavevmode {\epsfysize=7.5cm \epsffile{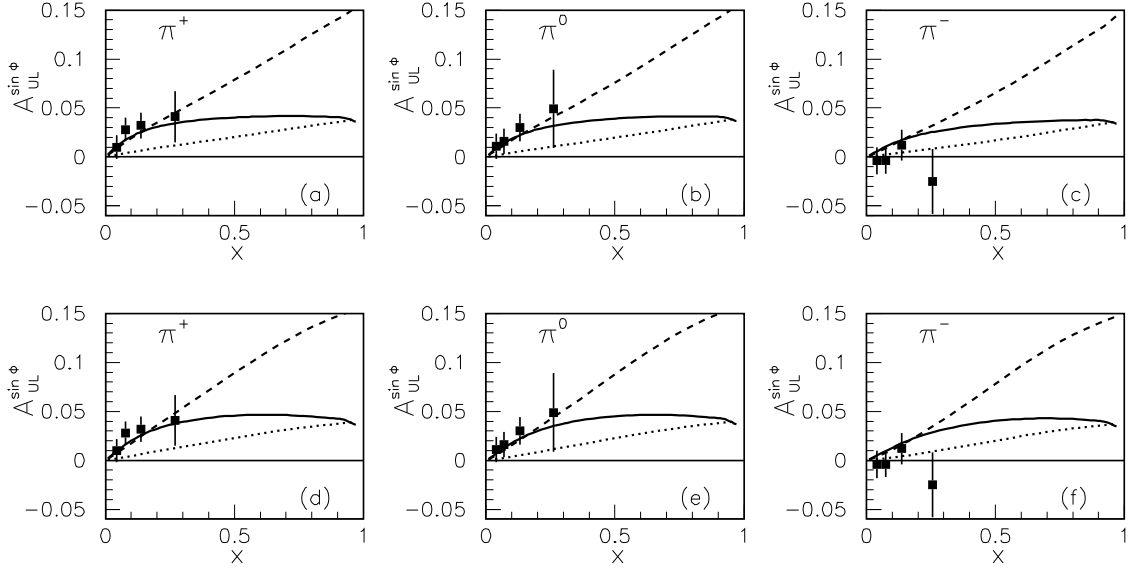}}
\end{center}
\caption[*]{\baselineskip 13pt The azimuthal asymmetries
$A_{UL}^{\sin \phi}$ for semi-inclusive $\pi^{+}$, $\pi^0$,  and
$\pi^-$ production in deep inelastic scattering of unpolarized
positron on the longitudinally polarized proton target. The upper
row corresponds to (a) $\pi^{+}$, (b) $\pi^{0}$, and (c) $\pi^-$
in the quark diquark model, and the lower row corresponds to
(d) $\pi^{+}$, (e) $\pi^{0}$, and (f) $\pi^-$ in the pQCD based
analysis. The dotted, dashed, and solid curves correspond to the
calculated results for Leading Approach, Approach 1, and Approach
2, respectively, of different assumptions of distribution functions
and fragmentation functions. The data are taken from the HERMES experiments
\cite{HERMES00,HERMES01}. }\label{msy10f1}
\end{figure}

We first check the difference of the azimuthal asymmetries between
different approaches of distribution functions and fragmentation
functions. In this situation we take into account the
contribution from both the favored and unfavored fragmentation
functions, i.e., with the new parametrization of both $D(z)$ and
$\hat{D}(z)$ in Eq.~(\ref{DD}) as inputs. The numerical results
for Leading Approach, Approach 1, and Approach 2 are given in
Fig.~\ref{msy10f1}. We find that both Approach 1 and Approach
2 differ significantly from Leading Approach, and this implies
that contribution from the distribution functions ${\tilde h}_L(x)$
and $h_{1L}^{\perp (1) }(x)$ are not negligible in the available
HERMES experimental region. We also found that the magnitude of
the azimuthal asymmetries from Leading Approach is below the
data, thus Leading Approach cannot describe the available
$\pi^+$ and $\pi^0$ data.
However, both Approach 1 and Approach 2
with the two model transversity distributions
are consistent with the available data, except for the data point
at $x=0.26$ for $\pi^-$ production. But the predictions from
Approach 1 and Approach 2 have a large difference at large $x$, and
we expect that further experimental data at large $x$ will be able to 
distinguish between the two approaches. We can also arrive at a following
conclusion:  although the HERMES azimuthal asymmetries for
longitudinally polarized target \cite{HERMES00,HERMES01} can
provide a rough estimate of the quark transversity distributions
by comparing theoretical calculations with the experimental data,
they do not make a direct measurement of the quark transversity
distributions since the contribution from other quark
distribution functions might be also sizable.

\vspace{0.3cm}

\subsection{Different cases of favored and unfavored fragmentation}

\begin{figure}[htb]
\begin{center}
\leavevmode {\epsfysize=7.5cm \epsffile{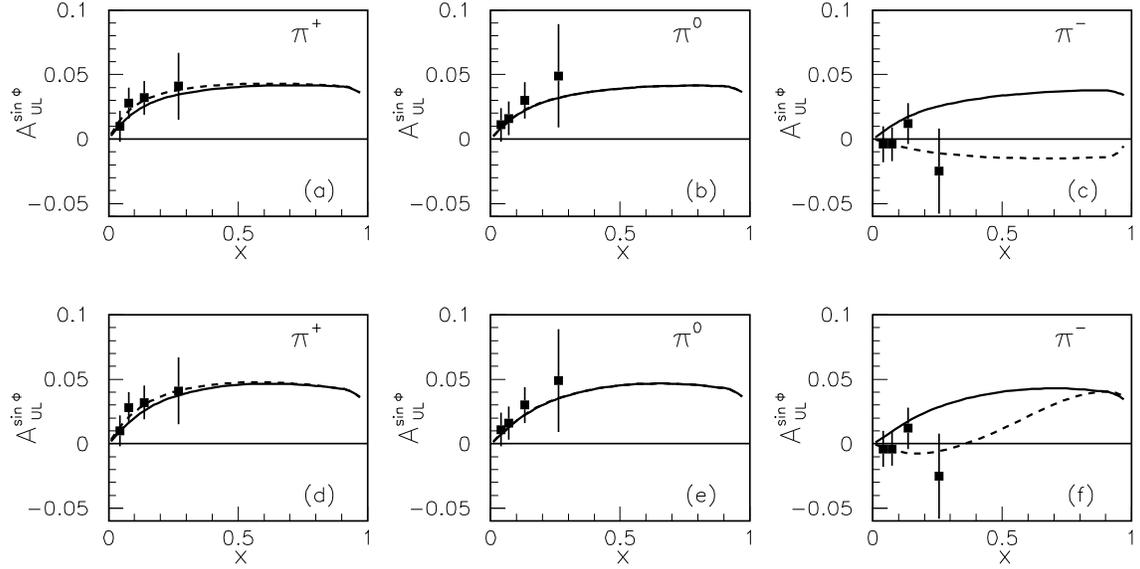}}
\end{center}
\caption[*]{\baselineskip 13pt The azimuthal asymmetries
$A_{UL}^{\sin \phi}$ for semi-inclusive $\pi^{+}$, $\pi^0$,  and
$\pi^-$ production in deep inelastic scattering of unpolarized
positron on the longitudinally polarized proton target. The upper
row corresponds to (a) $\pi^{+}$, (b) $\pi^{0}$, and (c) $\pi^-$
in the quark diquark model, and the lower row corresponds to
(d) $\pi^{+}$, (e) $\pi^{0}$, and (f) $\pi^-$ in the pQCD based
analysis. The solid and dashed curves correspond to the calculated
results for Case 1 with both favored and unfavored fragmentation
and Case 2 with only favored fragmentation, respectively.
The curves of the two cases overlap with each other for $\pi^0$
production.
}\label{msy10f2}
\end{figure}

We next check the contribution from the unfavored fragmentation
functions $\hat{D}(z)$. For this purpose we compare the
calculations in two cases: with both favored and unfavored
fragmentation functions as Case 1, and with only favored
fragmentation functions as Case 2. In order to simplify the
discussion, we will only perform calculations for Approach 2, and
the results are given in Fig.~\ref{msy10f2}. We find that there is
almost no difference between the calculated results in the two
cases for $\pi^+$ and $\pi^0$ production, and this supports the
favored fragmentation dominance assumption for the $\pi^+$ and
$\pi^0$ production, as was adopted in the literature.
However, the situation is quite different in
the two cases for $\pi^-$ production. This can be seen from
Fig.~\ref{msy10f2} (c) and (f), where the calculated results for
the two cases differ significantly. This implies that the
unfavored fragmentation functions have a large contribution to 
$\pi^-$ production. This can be easily understood, since the most
important unfavored process for $\pi^-$ production is $u \to
\pi^-$. The big difference between Case 1 and Case 2 shows that
the $u \to \pi^-$ process is not negligible due to the
$u$-dominance of the quark distributions in the proton target at large
$x$.

\vspace{0.3cm}

It is interesting to notice that, although there is clear evidence
for the non-zero azimuthal asymmetry for $\pi^+$ and $\pi^0$
production, the available data for $\pi^-$ production are
consistent with no azimuthal asymmetry. The data are consistent
with the calculated results with only favored fragmentation
functions, i.e., Case 2 \cite{Efr00,MSY9}.
However, the calculated results of Case 1 predict
non-zero azimuthal asymmetry for $\pi^-$ production in both of the
quark diquark model and the pQCD based analysis. From
Fig.~\ref{msy10f2} (c) and (f) we find that the data point for
$\pi^-$ production at $x=0.26$ seems to be in disagreement with
this prediction. We suggest more analysis on $\pi^-$ production
at the  $x>0.2$ region in order to test this prediction.

\begin{figure}[htb]
\begin{center}
\leavevmode {\epsfysize=7.5cm \epsffile{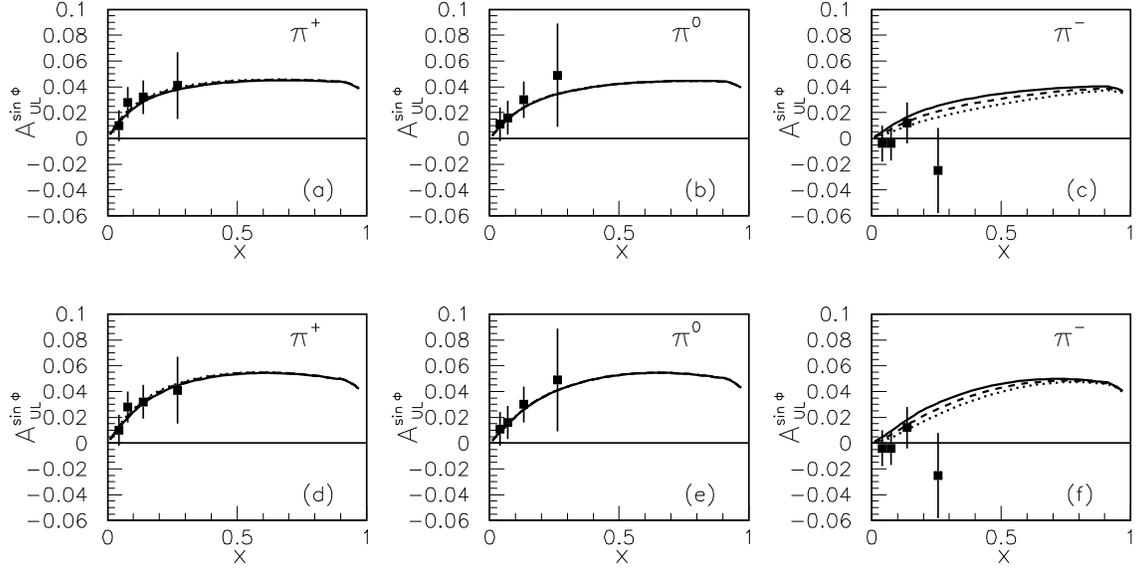}}
\end{center}
\caption[*]{\baselineskip 13pt The azimuthal asymmetries
$A_{UL}^{\sin \phi}$ for semi-inclusive $\pi^{+}$, $\pi^0$,  and
$\pi^-$ production in deep inelastic scattering of unpolarized
positron on the longitudinally polarized proton target. The upper
row corresponds to (a) $\pi^{+}$, (b) $\pi^{0}$, and (c) $\pi^-$
in the quark diquark model, and the lower row corresponds to
(d) $\pi^{+}$, (e) $\pi^{0}$, and (f) $\pi^-$ in the pQCD based
analysis. The solid, dashed, and dotted curves correspond to the
calculated results for three different options of unfavored
fragmentation functions with $n=2,3,4$, respectively. The cuts for
$z$ are $0.2<z<0.7$. }\label{msy10f3}
\end{figure}

\begin{figure}[htb]
\begin{center}
\leavevmode {\epsfysize=7.5cm \epsffile{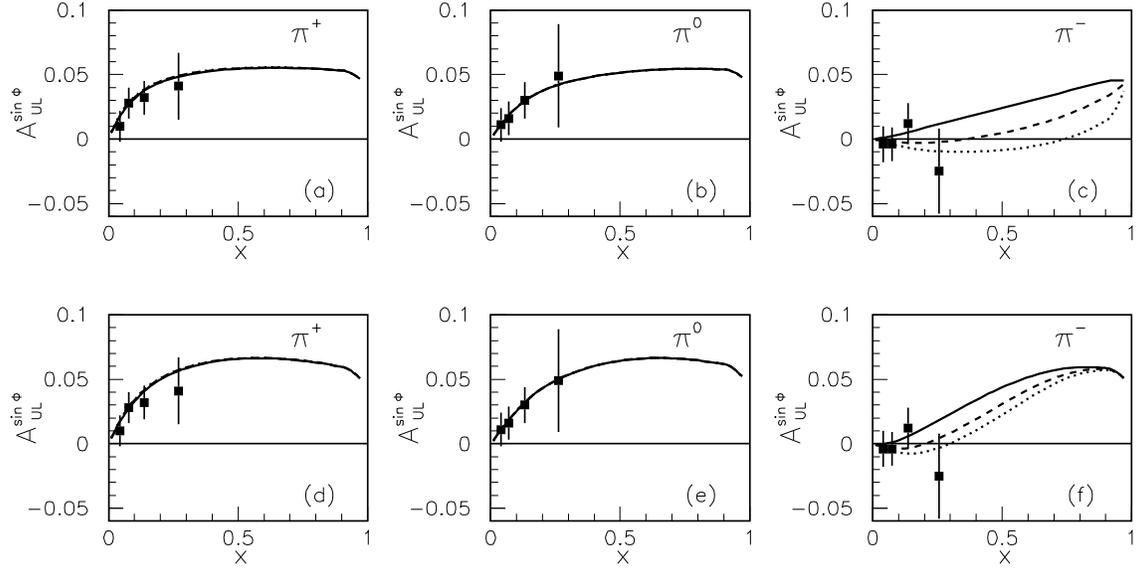}}
\end{center}
\caption[*]{\baselineskip 13pt The same as Fig.~\ref{msy10f3}, but with
cuts $0.6<z<0.8$.
}\label{msy10f4}
\end{figure}

We need to examine the possible ambiguities from the uncertainties of
the unfavored fragmentation functions. For this purpose we choose
the three options of $\hat{D}(z)=(1-z)^n D(z)$ with $n=2,3,4$,
respectively, and present the calculated results in
Fig.~\ref{msy10f3} with cuts $0.2 <z<0.7$, and in
Fig.~\ref{msy10f4} with cuts $0.6<z<0.8$, respectively. For the
available data with cuts $0.2 <z<0.7$, the three different options
of the unfavored fragmentation functions do not have big
difference, even for $\pi^-$ production, as shown in Fig.~\ref{msy10f3}. 
However, for the cuts
with higher $z$, where the unfavored fragmentation is more
suppressed than the favored ones for larger $n$, there is a large
difference between the calculated results for $\pi^-$ production
with the three options, as shown in Fig.~\ref{msy10f4}. 
Also there is a big difference between the
predictions for $\pi^-$ production with the two different models
of transversity distributions. Therefore the $\pi^-$ production at
large $x$ and large $z$ is sensitive to the uncertainties of the
unfavored fragmentation functions, and this is a region deserving
careful experimental studies.

\vspace{0.3cm}

\vspace{0.3cm}

With the contribution from only the favored fragmentation functions considered, it is found \cite{MSY9} that
there is a big difference between the azimuthal asymmetries for $\pi^-$ production predicted
by the quark diquark model and by the pQCD based analysis, as can be seen from the
dashed curves in Fig.~\ref{msy10f2} (c) and (f).
However, after taking into
account the unfavored fragmentation processes, we find a much reduced
difference between the two
different model predictions, see the solid curves
in Fig.~\ref{msy10f2} (c) and (f).
Thus a clear distinction between the two
different predictions will require very high precision measurement in the region with
both large $x$ and large $z$ for the proton target, and
this is difficulty to access experimentally. It also reminds us that the
$\pi^+$, $\pi^0$, and $\pi^-$ production from the proton target are mainly
controlled by the $u$ quark fragmentation, due to the $u$ quark dominance in the proton
target at large $x$. Thus the measured azimuthal asymmetries by the HERMES collaboration
\cite{HERMES00,HERMES01} are mainly controlled
by the $u$ quark transversity distribution $\delta u(x)$, and they are not sensitive
to the $d$ quark transversity distribution $\delta d(x)$.

\subsection{Deuteron and neutron targets}

\begin{figure}[htb]
\begin{center}
\leavevmode {\epsfysize=7.5cm \epsffile{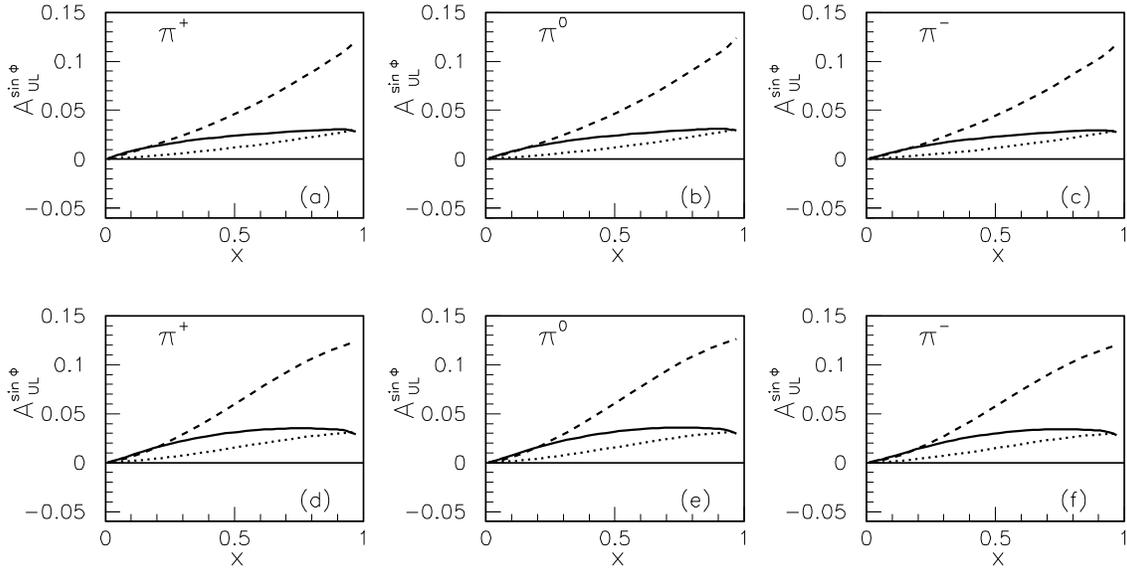}}
\end{center}
\caption[*]{\baselineskip 13pt The azimuthal asymmetries
$A_{UL}^{\sin \phi}$ for semi-inclusive $\pi^{+}$, $\pi^0$,  and
$\pi^-$ production in deep inelastic scattering of unpolarized
positron on the longitudinally polarized deuteron target. The
upper row corresponds to (a) $\pi^{+}$, (b) $\pi^{0}$, and (c)
$\pi^-$ in the quark diquark model, and the lower row
corresponds to (d) $\pi^{+}$, (e) $\pi^{0}$, and (f) $\pi^-$ in
the pQCD based analysis.  The dotted, dashed, and solid curves
correspond to the calculated results for Leading Approach,
Approach 1, and Approach 2, respectively, of different assumptions
of quark distributions and fragmentation functions.
}\label{msy10f5}
\end{figure}

We now consider the possibility to distinguish between
different models by using the deuteron as target, where there are equal numbers
of $u$ and $d$ quarks from isopin symmetry.
To calculate the azimuthal asymmetries for the deuteron target, we still
use the formalism
in Sec.II, with the target polarization $S_D=0.75$ multiplied by an correction factor
\begin{equation}
\left(1-\frac{3}{2} \omega_{D}\right),
\end{equation}
where $\omega_{D}=0.05\pm 0.01$ is the probability of the deuteron
to be in the $D$-state. The azimutahl asymmetries for $\pi^+$,
$\pi^0$, and $\pi^-$ production are calculated for different
approaches of distribution functions and fragmentation functions,
and with the quark transversity distributions of the quark
diquark model and the pQCD based analysis, respectively. For the
unpolarized fragmentation functions, we use the new parametrization
Eq.~(\ref{DD}), and for the kinematical cuts  we use the same as
those in the HERMES experiments. The results are given in
Fig.~\ref{msy10f5}. From there we find that the deuteron is also not very
sensitive to the two different sets of transversity distributions.

\vspace{0.3cm}

\vspace{0.3cm}
\begin{figure}[htb]
\begin{center}
\leavevmode {\epsfysize=7.5cm \epsffile{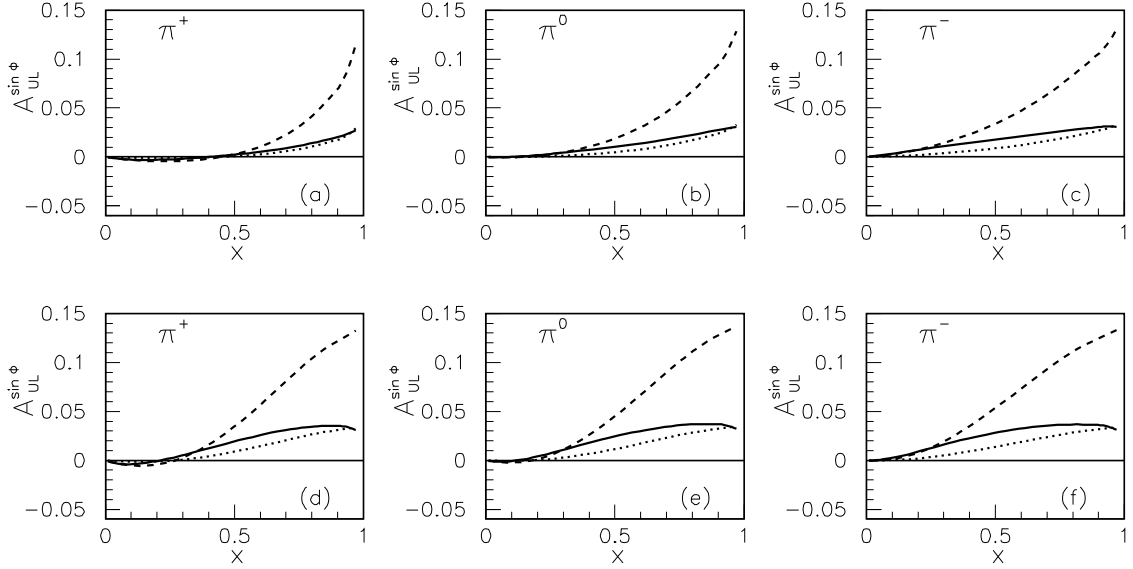}}
\end{center}
\caption[*]{\baselineskip 13pt The same as Fig.~\ref{msy10f5}, but with
the neutron target.
}\label{msy10f6}
\end{figure}

We now check whether the situation can be improved if we use the
neutron as target. This can be experimentally accessed by
measuring the numerators and denominators of the spin asymmetries
(\ref{SA1phi}) from proton and deuteron targets respectively, and
then extract the numerator and denominator for the neutron contribution
to calculate the spin symmetry of the neutron. The calculated results
are given in Fig.~\ref{msy10f6}. We find that there is a somewhat  bigger
difference between the predictions of the quark diquark model and
the pQCD based analysis. In order to understand this behavior,
we consider the $\pi^+$ production for a neutron target, shown in
Fig.~\ref{msy10f6} (a) and (d), and compare them with the $\pi^-$
production for a proton target, shown in Fig.~\ref{msy10f1} (c)
and (f). For the proton case, the favored fragmentation process is
$d \to \pi^-$, while the unfavored process is $u \to \pi^-$. The
unfavored process is important due to the dominance of $u$ quark
distribution over $d$ inside the proton target at large $x$. For the neutron case, the
favored fragmentation process is $u \to \pi^+$, while the
unfavored process is $d \to \pi^+$. The unfavored process is
suppressed by a factor of 4 (from the squared charge) compared to
the corresponding process of the proton target, while the favored
process is enhanced by a factor of 4 compared to the corresponding
process of the proton target. Thus the favored process $u \to
\pi^+$ relative to the unfavored process $d \to \pi^+$ for the
neutron target is enhanced by a factor of 16 compared to the
corresponding favored process $d \to \pi^-$ relative to the
unfavored process $u \to \pi^-$ for the proton. This implied that the favored
process of $\pi^+$ production for the neutron target plays a more
important role than the corresponding process of $\pi^-$
production for the proton target, where the unfavored process is
more important and cannot be neglected. Thus we suggest to
distinguish between the quark diquark model and the pQCD based
analysis by measuring  $\pi^+$ production from a neutron
target. The $\pi^+$ production for a neutron target is
sensitive to the $u$ quark transversity distribution of the neutron, which is
essentially the $d$ quark transversity distribution of the proton.
Thus combination of pion production from both the proton and
neutron targets can measure the flavor structure of the quark
transversity distributions.

\subsection{The azimuthal asymmetries of $A_{UL}^{\sin 2 \phi }$}

\begin{figure}[htb]
\begin{center}
\leavevmode {\epsfysize=7.5cm \epsffile{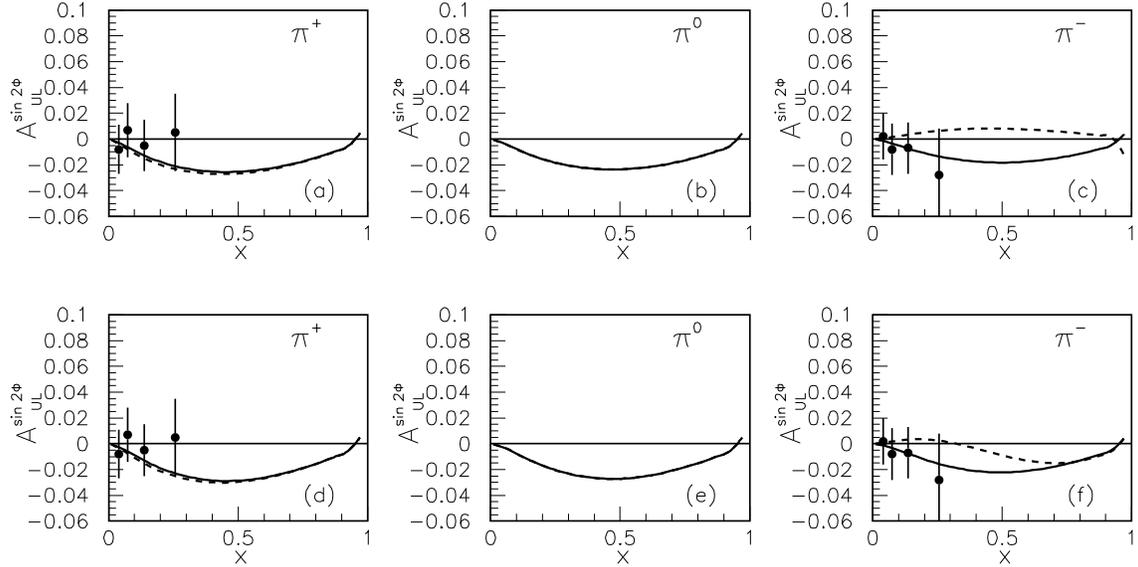}}
\end{center}
\caption[*]{\baselineskip 13pt The azimuthal asymmetries
$A_{UL}^{\sin 2 \phi}$ for semi-inclusive $\pi^{+}$, $\pi^0$,  and
$\pi^-$ production in deep inelastic scattering of unpolarized
positron on the longitudinally polarized {\it proton} target (with
polarization $S=0.86$). The upper row corresponds to (a)
$\pi^{+}$, (b) $\pi^{0}$, and (c) $\pi^-$ in the quark diquark
model, and the lower row corresponds to (d) $\pi^{+}$, (e)
$\pi^{0}$, and (f) $\pi^-$ in the pQCD based analysis. The solid
and dashed curves correspond to the calculated results for Case 1
with both favored and unfavored fragmentation and Case 2 with only
favored fragmentation, respectively. }\label{msy10f7}
\end{figure}

\begin{figure}[htb]
\begin{center}
\leavevmode {\epsfysize=7.5cm \epsffile{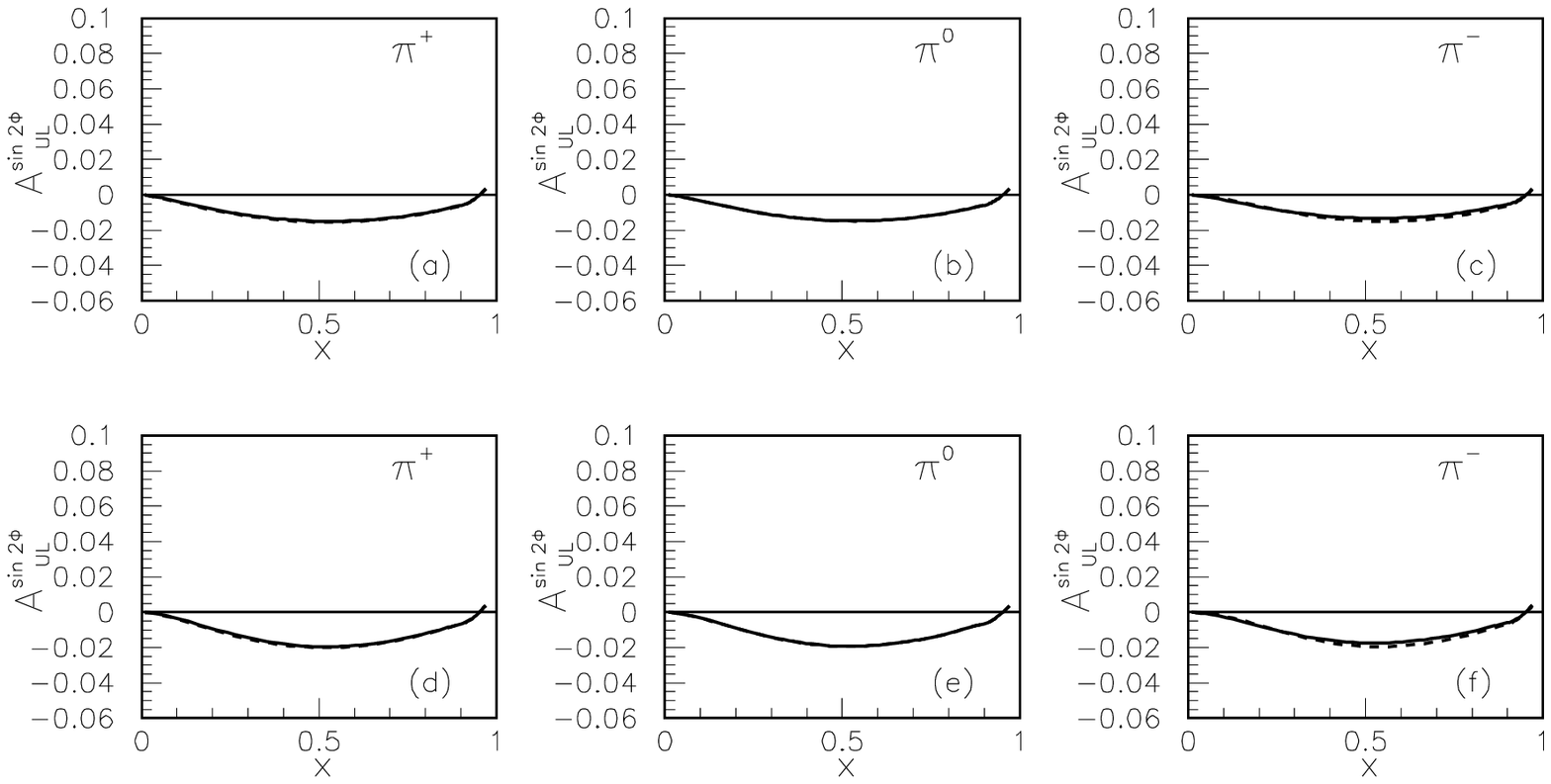}}
\end{center}
\caption[*]{\baselineskip 13pt The same as Fig.~\ref{msy10f7}, but for
the {\it deuteron} target with polarization $S=0.75$. }\label{msy10f8}
\end{figure}

\begin{figure}[htb]
\begin{center}
\leavevmode {\epsfysize=7.5cm \epsffile{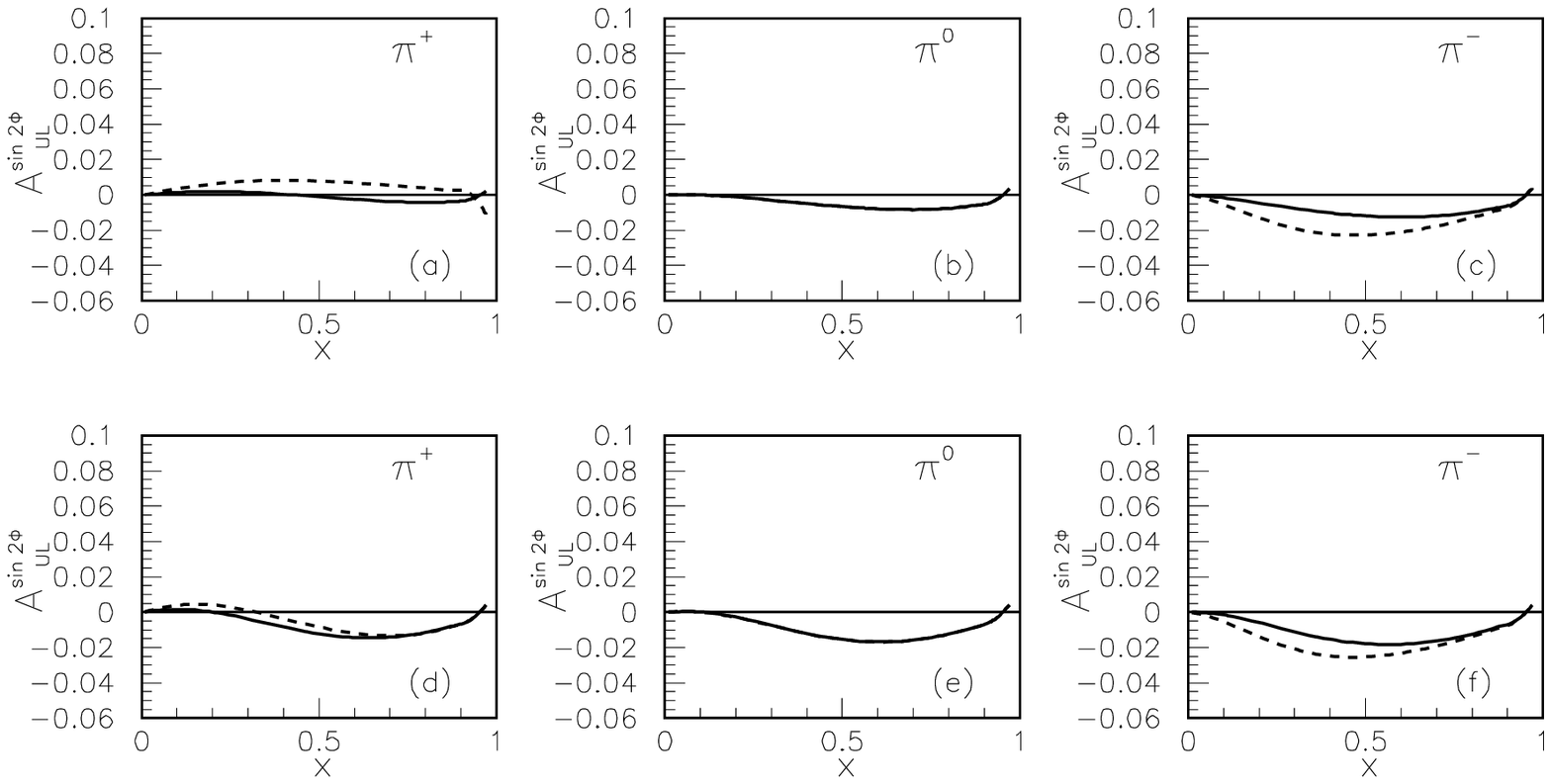}}
\end{center}
\caption[*]{\baselineskip 13pt The same as Fig.~\ref{msy10f7}, but for
the {\it neutron} target with polarization $S=0.75$.   }\label{msy10f9}
\end{figure}

For the azimuthal asymmetries with weighting function $\sin
2\phi$, i.e., $A_{UL}^{\sin 2 \phi }$, there is only one quark
distribution function $h_{1L}^{\perp (1) }(x)$ involved in the
numerator of (\ref{SA2phi}). In both the Leading
Approach and Approach 1, this distribution function is assumed to
be zero. Therefore the azimuthal asymmetries $A_{UL}^{\sin 2 \phi
}$ can only be calculated in Approach 2, and we present the
results for the proton, deuteron, and neutron targets in
Figs.~\ref{msy10f7}, \ref{msy10f8}, and \ref{msy10f9},
respectively, in the two different models of quark transversity
distributions. We find that the magnitude of these asymmetries are
not so small in the two models, thus it is possible to 
measure them with high precision measurements. In the case of a
proton target, the calculated results are similar for the two
models with the available HERMES kinematical cuts, and they are
both compatible with the available HERMES data. The calculated
results are also similar in the two models in the case of the
deteron target. However, larger differences between the predictions of 
the two models is found in the case of $\pi^+$ production from
a neutron target. Thus
precision measurement of $A_{UL}^{\sin 2 \phi }$ from  nucleon
targets may provide us with information on the distribution
function $h_{1L}^{\perp (1) }(x)$ and its flavor structure.

\vspace{0.3cm}

\vspace{0.3cm}

\vspace{0.3cm}

\section{Prediction of Pion Azimuthal Asymmetries for Transversely Polarized Target}

From the above discussions, we find that although the azimuthal
asymmetries off longitudinally polarized target may serve as a
measurement of the transversity distributions, there are still
some contribution from several unmeasured distribution functions
and fragmentation functions. Their contribution are not
negligible in the HERMES kinematical region, and also there are
some ambiguities concerning the size of their contribution from
different assumptions, as we can see from the different
predictions between Approach 1 and Approach 2 in the last section.
Therefore it is necessary to look for more clean processes for a
direct measurement of the quark transversity. In fact, it has been known that the azimuthal
asymmetries off transversely polarized target are directly
connected to the Leading Approach term we discussed for the
longitudinally polarized situation. There has been some
preliminary results from the SMC experiment \cite{SMC99}, indicating
evidence of non-zero azimuthal asymmetry for pion production from
a transversely polarized target. Also the HERMES collaboration is
planning to measure the azimuthal asymmetries of pion production
from transversely polarized targets  in the near future \cite{Kor00}. It is
thus necessary to make clear predictions on the quantities they
will measure by using the two sets of transversity distributions
given by the quark diquark model and the pQCD based analysis.

\begin{figure}[htb]
\begin{center}
\leavevmode {\epsfysize=7.5cm \epsffile{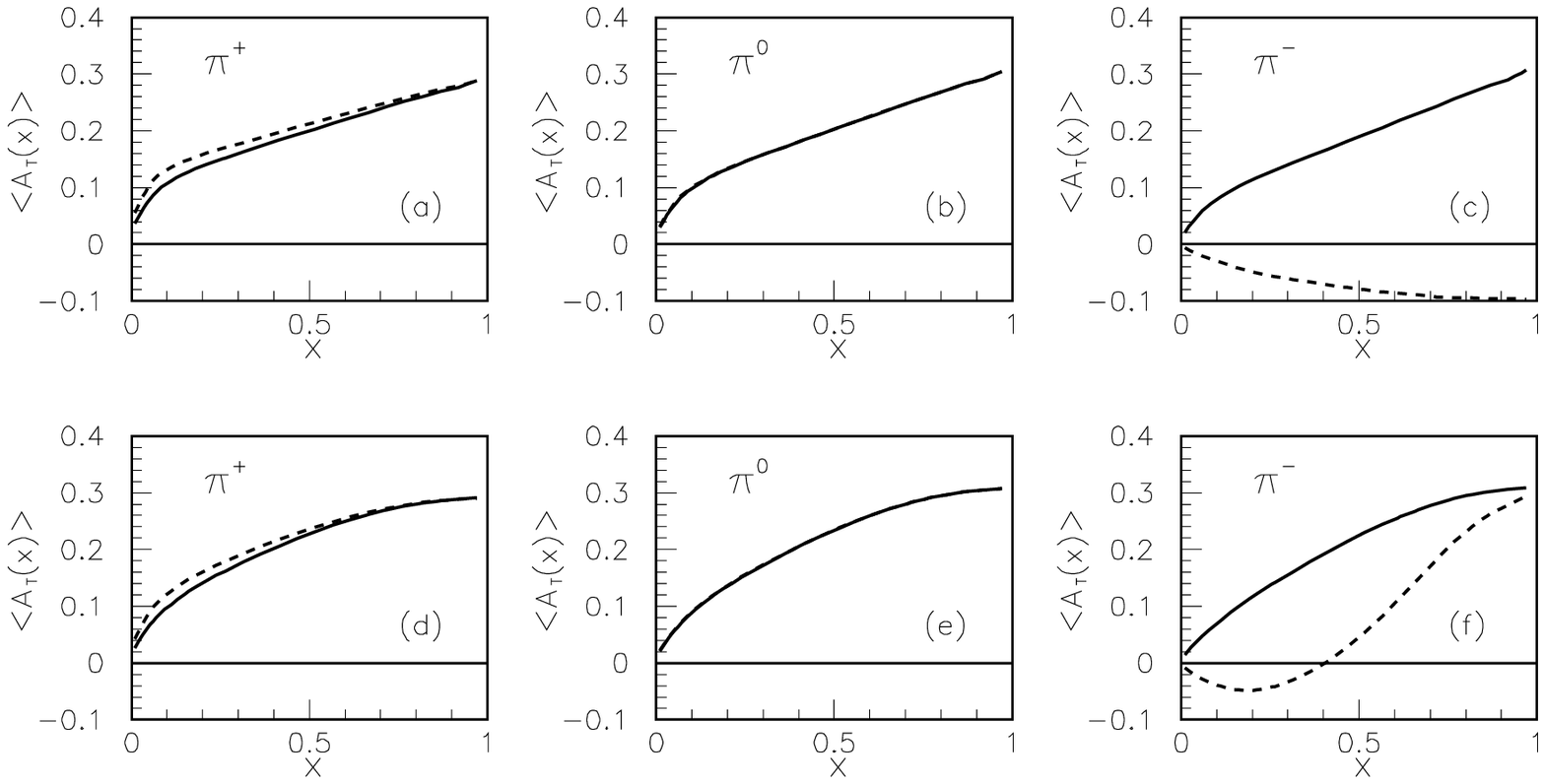}}
\end{center}
\caption[*]{\baselineskip 13pt The azimuthal asymmetries
$\left<A_T(x) \right>$ for semi-inclusive $\pi^{+}$, $\pi^0$,  and
$\pi^-$ production in deep inelastic scattering of unpolarized
positron on the transversely polarized {\it proton} target. 
The upper row corresponds to (a)
$\pi^{+}$, (b) $\pi^{0}$, and (c) $\pi^-$ in the quark diquark
model, and the lower row corresponds to (d) $\pi^{+}$, (e)
$\pi^{0}$, and (f) $\pi^-$ in the pQCD based analysis. The solid
and dashed curves correspond to the calculated results for Case 1
with both favored and unfavored fragmentation and Case 2 with only
favored fragmentation, respectively. }\label{msy10f10}
\end{figure}

\begin{figure}[htb]
\begin{center}
\leavevmode {\epsfysize=7.5cm \epsffile{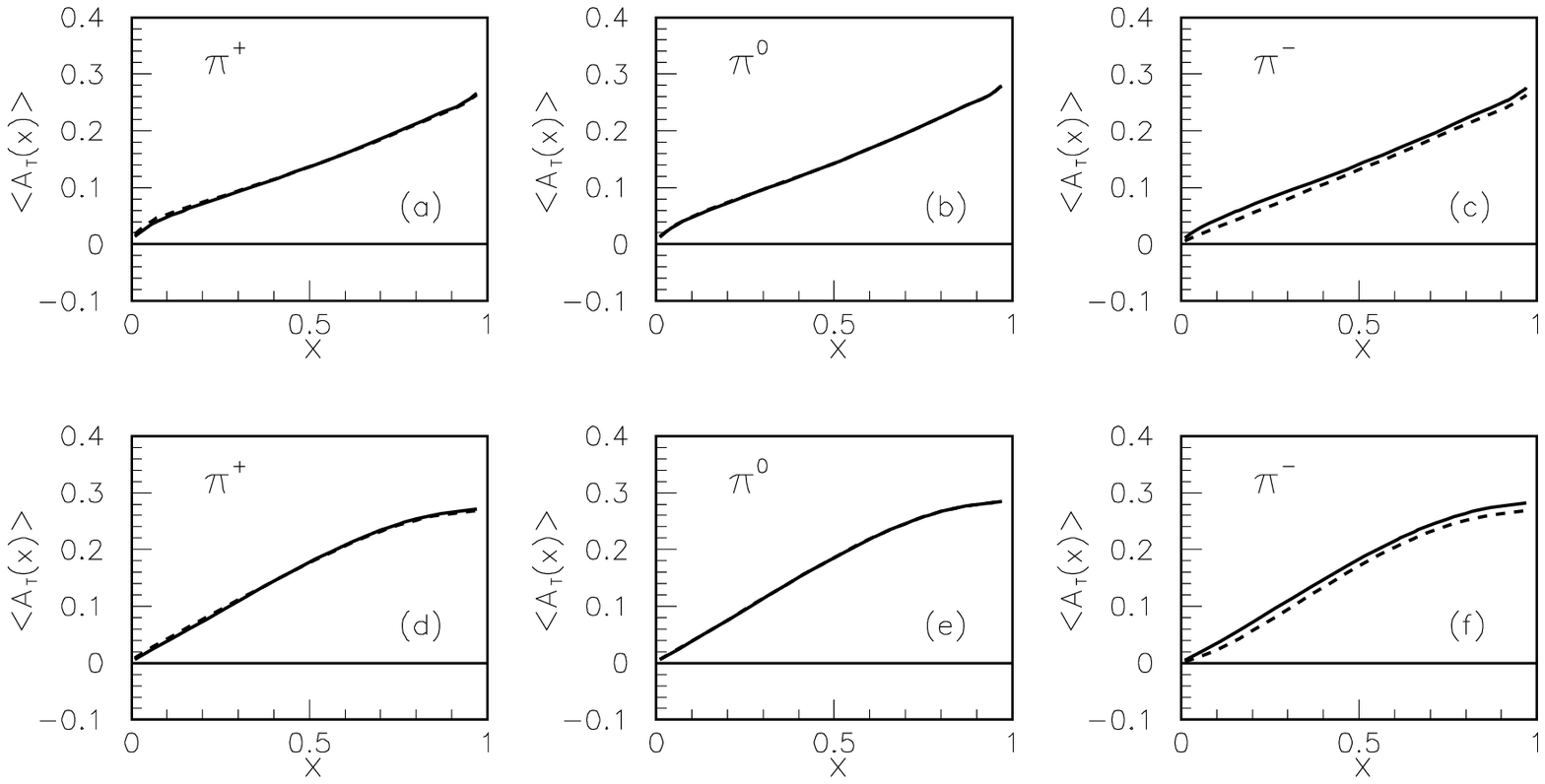}}
\end{center}
\caption[*]{\baselineskip 13pt The same as Fig.~\ref{msy10f10}, but for
the {\it deuteron} target.
}\label{msy10f11}
\end{figure}

\begin{figure}[htb]
\begin{center}
\leavevmode {\epsfysize=7.5cm \epsffile{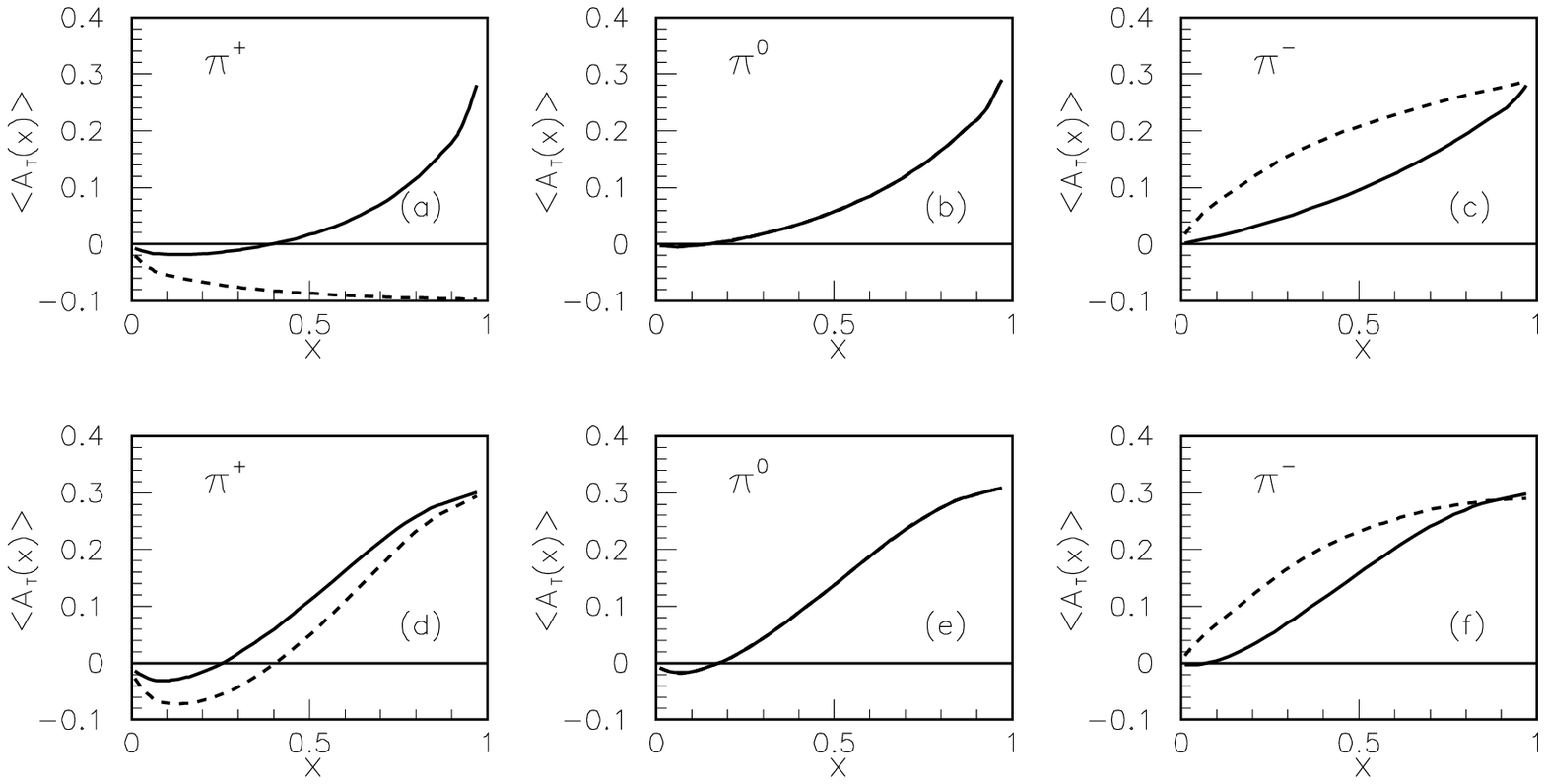}}
\end{center}
\caption[*]{\baselineskip 13pt The same as Fig.~\ref{msy10f10}, but for
the {\it neutron} target.  }\label{msy10f12}
\end{figure}

More specifically, for an unpolarized lepton beam and a transversely polarized target,
the following weighted asymmetry provides access to the quark transversity distribution
\cite{Kot97}
\begin{equation}
A_T(x,y,z)=
\frac{\int \d \phi^{l}\int \d^2 P_{h\perp} \frac{|P_{h\perp}|}{z M_h} \sin (\phi_s^l+\phi_h^l)
\left(\d \sigma^{\uparrow}-\d \sigma^{\downarrow} \right)}
{\int\d \phi^l \int \d^2 P_{h\perp} \left(\d \sigma^{\uparrow}+\d \sigma^{\downarrow} \right)},
\label{AT}
\end{equation}
where $\uparrow$ ($\downarrow$) denotes target up (down) transverse polarization.
The azimuthal angles are defined in the transverse space giving the orientation
of the lepton plane ($\phi^l$) and the orientation of the hadron plane ($\phi_h^l=\phi_h-\phi^l$)
or spin vector ($\phi^l_s=\phi_s-\phi^l$) with respect to the lepton plane. The asymmetry
(\ref{AT}) can be calculated from the distribution functions and fragmentation functions by
\cite{Kot97}
\begin{equation}
A_T(x,y,z)=S_T \frac{(1-y) \sum_q e^2_q \delta q(x) H_1^{\perp (1) q}(z)}{(1-y+y^2/2)\sum_q e^2_q q(x) D_1^q(z)},
\end{equation}
where $S_T$ is the target polarization. Assuming $S_T=0.75$ and
the same kinematical cuts as in the MERMES experiments, and with
all of the quantities discussed as before, we can calculate the
asymmetry $A_T$ with the two different sets of quark transversity
distributions. The integration over $y$ and $z$ should be
performed for both the numerator and denominator. The predictions for 
a proton target are given in Fig.\ref{msy10f10},
from where
we notice that the azimuthal asymmetries of $\pi^{+}$, $\pi^0$, and
$\pi^-$ production from a proton target are mainly controlled by
the $u$ quark transversity distribution, and they can serve to
measure $\delta u(x)$. The predictions for pion production from a
deuteron target, given in Fig.~\ref{msy10f11},
are also found to have small differences between the 
two different models. However, the azimuthal asymmetries from a
neutron target
are predicted to have a large difference between
the quark diquark model and the pQCD based analysis, as shown in
Fig.~\ref{msy10f12}. Thus they can
serve to measure the flavor structure of quark transversity
distributions, i.e., one can measure both $\delta u(x)$ and $\delta
d(x)$ by using the azimuthal asymmetries from both  proton and
neutron targets.

\vspace{0.3cm}

\vspace{0.3cm}

\vspace{0.3cm}

\section{Summary and Conclusions}

In this paper, we analyzed in detail the azimuthal spin
asymmetries of pion production, in semi-inclusive deep-inelastic
scattering of unpolarized charged lepton beams on longitudinally
and transversely polarized nucleon targets. Various assumptions
and approximations of the distribution functions and fragmentation
functions were examined. It was found that different approaches of
distribution functions and fragmentation functions lead to
different predictions of the azimuthal asymmetries in the
available HERMES kinematical region. This means that the
effects from unknown distribution functions may have sizable
effects on the azimuthal asymmetries of the HERMES experiments,
thus one needs to consider them before attributing the
HERMES data as a measurement of transversity distributions. We
also found that the predictions of $\pi^-$ production from the
proton target are quite different for two cases of both favored
and unfavored fragmentation, and of only favored fragmentation.
This means that the unfavored fragmentation functions play an
important role for  $\pi^-$ production from a proton target,
due to the dominance of $u$ quarks inside the proton target, thus
the unfavored $u \to \pi^-$ fragmentation are not negligible. This
point was not considered in the available studies of azimuthal
asymmetries. The pion production from the proton target is most
suitable to study the $u$ quark transversity distribution. In
combination with the pion production from a neutron target, it
is possible to measure both of the $u$ and $d$ quark transversity
distributions. Thus we suggest to measure the azimuthal symmetries
of pion production from the neutron target, in order to test
different model predictions. Predictions of azimuthal asymmetries
of pion production in semi-inclusive deep-inelastic scattering of
unpolarized charged lepton beams on  transversely polarized
nucleon targets are also
given, and different predictions between two different models are
given for pion production from a neutron target. This study
will be useful for designing  experiments aiming at measuring
transversity distributions through pion production in
semi-inclusive deep-inelastic scattering of unpolarized beams on
longitudinally and transversely polarized targets. It is also
useful for theoretical studies aiming at extracting transversity
distributions from experimental observation of azimuthal
asymmetries of pion production in semi-inclusive deep-inelastic
scattering.

{\bf Acknowledgments: }
This work is
partially supported by National Natural Science Foundation of
China under Grant Numbers 19975052, 19875024, and 10025523, by Fondecyt
(Chile) postdoctoral fellowship 3990048, and by Fondecyt (Chile) grant 8000017.

\end{document}